\documentclass[conference]{IEEEtran}
\IEEEoverridecommandlockouts
\usepackage{cite}
\usepackage{amsmath,amssymb,amsfonts}
\usepackage{algorithmic}
\usepackage{graphicx}
\usepackage{textcomp}
\usepackage{xcolor}
\usepackage{caption}
\usepackage{subcaption}
\usepackage{multirow}
\usepackage[flushleft]{threeparttable}
\def\BibTeX{{\rm B\kern-.05em{\sc i\kern-.025em b}\kern-.08em
    T\kern-.1667em\lower.7ex\hbox{E}\kern-.125emX}}
\usepackage{array}
\newcolumntype{C}[1]{>{\centering\arraybackslash}m{#1}}
\renewcommand{\baselinestretch}{0.90}

\begin{document}

\title{Enhancements for 5G NR PRACH Reception: An AI/ML Approach}

\author{\IEEEauthorblockN{Rohit Singh\IEEEauthorrefmark{1},
Anil Kumar Yerrapragada\IEEEauthorrefmark{2},
Jeeva Keshav S\IEEEauthorrefmark{3},
Radha Krishna Ganti\IEEEauthorrefmark{4}}
\IEEEauthorblockA{Department of Electrical Engineering\\
Indian Institute of Technology
Madras \\ Chennai, India  600036\\
Email: \IEEEauthorrefmark{1}rohitsingh@smail.iitm.ac.in,
        \IEEEauthorrefmark{2}anilkumar@5gtbiitm.in,
        \IEEEauthorrefmark{3}jeevakeshavs@smail.iitm.ac.in,
	\IEEEauthorrefmark{4}rganti@ee.iitm.ac.in
}
}

\maketitle

\begin{abstract}
Random Access is an important step in enabling the initial attachment of a User Equipment (UE) to a Base Station (gNB). The UE identifies itself by embedding a Preamble Index (RAPID) in the phase rotation of a known base sequence, which it transmits on the Physical Random Access Channel (PRACH). The signal on the PRACH also enables the estimation of propagation delay, often known as Timing Advance (TA), which is induced by virtue of the UE's position. Traditional receivers estimate the RAPID and TA using correlation-based techniques. This paper presents an alternative receiver approach that uses AI/ML models, wherein two neural networks are proposed, one for the RAPID and one for the TA. Different from other works, these two models can run in parallel as opposed to sequentially. Experiments with both simulated data and over-the-air hardware captures highlight the improved performance of the proposed AI/ML-based techniques compared to conventional correlation methods.  
\end{abstract}


\section{Introduction}
In any cellular system, the UEs have to first identify and attach themselves to a base station. This procedure is known as Random Access (RA). In order to facilitate this procedure, the 5G standards have a provision for a dedicated channel known as the Physical Random Access Channel (PRACH). On the PRACH, the UE sends a randomly rotated version of a known base sequence. The identity of the UE, known as Random Access Preamble ID (RAPID), is embedded in this rotation. Typically the base sequence is a Zadoff-Chu sequence~\cite{1057798}, well known for its Constant Amplitude Zero Auto Correlation (CAZAC) properties. The receiver can then leverage these properties and extract the RAPID by correlation with the known base sequence. In addition,  the correlation also helps in the estimation of the propagation delay, which the signal incurs as it traverses the distance between the UE and the gNB. The delay value is reported back to the UE and is then used by the UE to advance its future transmissions. In this paper, we refer to the propagation delay value as the Timing Advance (TA). Though the primary interest lies in estimating the TA, the propagation delay estimate also captures any fine drift in time synchronization between the UE and gNB. 

In ideal conditions, the correlation leads to a single peak that contains information about the RAPID and the Timing Advance. Traditional receivers use thresholds in order to identify the correct peak. However, setting up these thresholds requires tedious experiments and simulations~\cite{hu2012method}. Improper thresholds result in false alarms and/or missed detections. Furthermore, since the RAPID selection is random, it is possible for multiple UEs to use the same RAPID, possibly leading to collisions. Collisions lead to re-transmissions, which in turn increases the initial attach latency and also consumes more energy for every re-transmission.

In this paper, we direct our attention toward the RAPID detection and TA estimation of a single user, both of which we recast as AI/ML classification problems. We make the following contributions.
\subsection{Our Contributions}
\begin{itemize}
    \item Prior AI/ML implementations of the PRACH receiver rely upon input data from multiple correlation zones ~\cite{quek}. This paper presents an approach that avoids correlations altogether. Both the RAPID and TA classification models require no pre-processing of the input and take frequency domain data directly from the OFDM grid.
    
    \item Several existing works implement RAPID detection and TA estimation in a sequential manner, in which incorrect detection of RAPID could result in TA estimation errors. In the proposed receiver, the RAPID detection and TA estimation models can work in parallel using the same received signal input and are independent of each other. 

    \item In this paper, we study the effects of training and testing with different channel models. In particular, we study the generalization performance of the models under various configurations of delay spread values (TDLC150 and TDLC300).
     
    \item Most works in AI/ML for wireless communications deal with simulated data and environments. Hardware impairments are rarely incorporated. In this paper, we analyze the performance of the proposed AI/ML models using Over-the-air (OTA) captures obtained from the 5G Testbed at IIT Madras~\cite{5gtbiitm}.

    
\end{itemize}

\section{Background on 5G NR Physical Random Access Channel}
This section offers background information on Random Access Procedures. We also describe, compare, and contrast existing prior works on both conventional and AI/ML-based receivers for PRACH.

\subsection{5G Random Access Procedures}
\begin{figure}
    \centering
    \includegraphics[width = 0.6\linewidth]{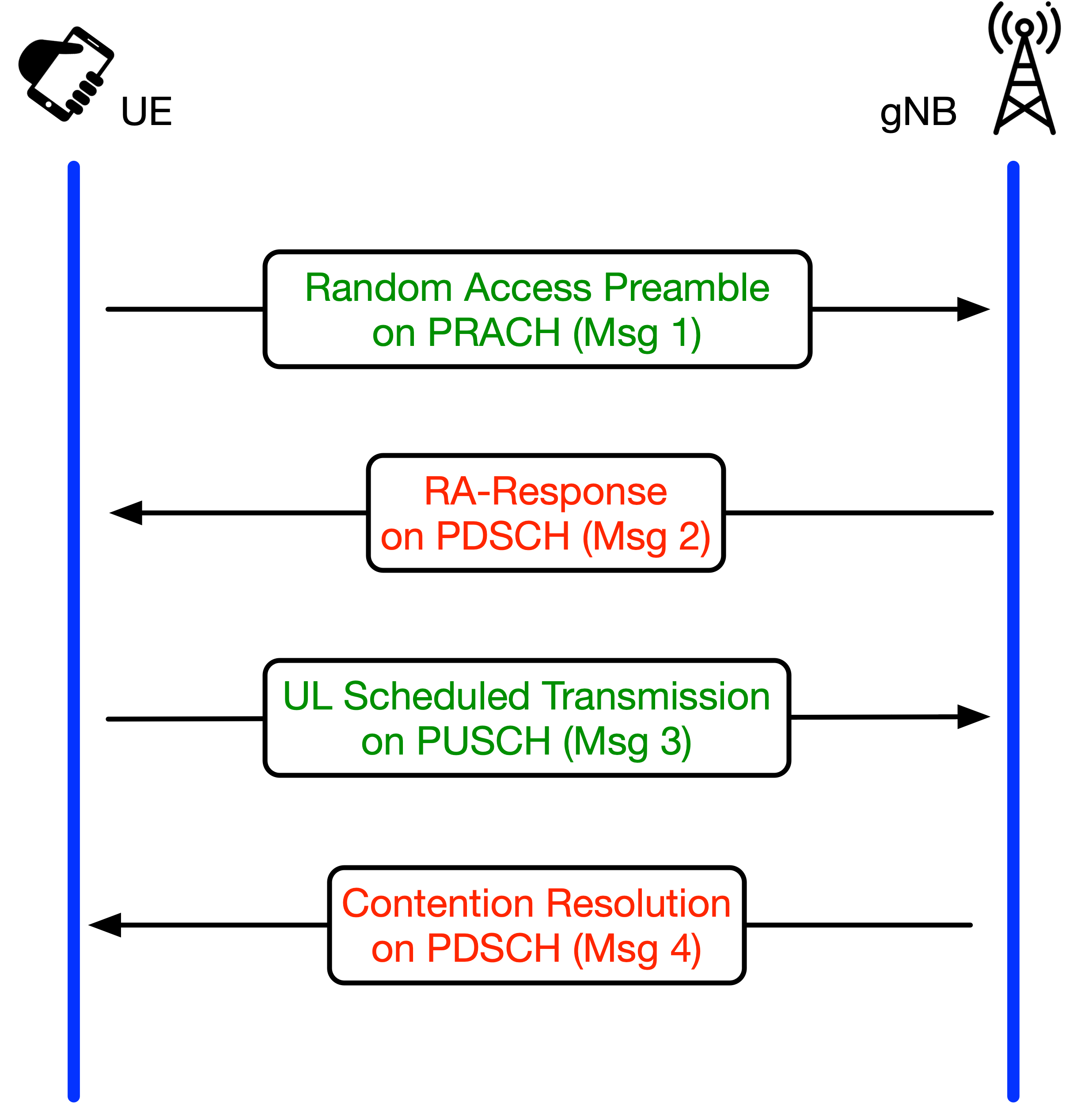}
    \caption{Contention Based Random Access (CBRA) Procedure}
    \label{fig: RA_Procedure}
\end{figure}

The 5G specification supports two types of Random Access (RA) procedures. The first is contention-based in which a UE is free to select any preamble randomly from a pool of preambles. The second is contention-free in which the gNB assigns preamble IDs to the UE. In this paper, we focus on the Contention Based Random Access (CBRA) procedure which is a 4-step process described in Fig.~\ref{fig: RA_Procedure}. The RA procedure begins with PRACH (Msg 1) signaling. In this step, the UE selects a RAPID from a pool of possible values and applies the corresponding cyclic shift to the base sequence as specified in~\cite{3gpp_38_211}. We note that at this point, the UE has already ascertained the PRACH time-frequency locations in the resource grid from prior messages in the Synchronization Signal Block (SSB) and System Information Block \#1 (SIB1), both of which are periodically broadcast by the gNB.

The gNB acknowledges decoding of the PRACH signal through a Random Access Response (RAR, Msg 2) on the Physical Downlink Shared Channel (PDSCH). This is immediately followed by a scheduled uplink transmission (Msg 3) on the Physical Uplink Shared Channel (PUSCH). The fourth step includes scheduled downlink transmissions (Msg 4) on the PDSCH for Contention Resolution (CR) and for further initial attachment processing that ends in the UE latching on to the gNB. If the PRACH detection is incorrect, it will be evident at Msg 3, and the RA process restarts, causing significant latency in the initial attach process. In order to reduce this initial attach latency, PRACH receivers should be robust enough to reduce incorrect detections even in multipath and low SNR environments.

\subsection{PRACH Sequence Generation and Transmission}
In this paper, we focus on the PRACH signaling in Msg1 of the RA process. PRACH sequence generation starts with the selection of a base sequence which is given by, 
\begin{equation}
    x_u(n) = e^{\frac{-j\pi un(n+1)}{L_{RA}}}, n = 0,1,....., L_{RA} -1.
    \label{eq: prach_zc}
\end{equation}
The length of the sequence $L_{RA}$, depends on the PRACH preamble format given by Table 6.3.3.1-2 in TS 38.211~\cite{3gpp_38_211}. The initial sequence number $u$ is selected from Table 6.3.3.1-4 in TS 38.211. A cyclic rotation is applied to $x_{u}(n)$ resulting in $x_{u,v}(n) = x_u((n + C_v) \text{ mod } L_{RA})$. The parameter $C_{v}$ is a function of the parameter $N_{CS}$ which is given in Table 6.3.3.1-7 in TS 38.211. The table indices are configured by higher layers. 

Without loss of generality, we use commonly configured parameters for the purpose of illustration and experiments. We use the following values: $L_{RA} = 139$ and $N_{CS} = 13$ resulting in $C_{v} = vN_{CS}$ where $v = 0,1,....., \left\lfloor \frac{L_{RA}}{N_{CS}} \right\rfloor-1$. 

The 5G standards have a provision for 64 Random Access Preamble Indices (RAPID). For the configured values, $\left\lfloor \frac{L_{RA}}{N_{CS}}\right\rfloor = 10$, and  hence each base sequence can accommodate $10$ RAPID values. In order to encompass $64$ RAPID values, $7$ base sequences are required. In this paper, we develop neural network models for a single base sequence and $10$ RAPID values. Other base sequences can be incorporated similarly.

The sequences $x_u(n)$ and $x_{u, v}(n)$ are generated in the time domain. In order to map $x_{u, v}(n)$ to the OFDM resource grid, it is first converted to the frequency domain as follows, 
\begin{equation}
    \begin{split}
    y_{u,v}(k) &= \sum_{n=1}^{L_{RA} - 1} x_{u,v}(n) e^{\frac{-j2\pi nk }{L_{RA}}} = X_u(k) e^{\frac{-j2\pi k C_v }{L_{RA}}},
    \end{split}
    \label{eq: prach_dft}
\end{equation}
where,
\begin{equation}
    X_u(k) = \sum_{n=1}^{L_{RA} - 1} x_{u}(n) e^{\frac{-j2\pi nk }{L_{RA}}}.
\end{equation}
Then, $y_{u, v}(k)$ can be mapped to the frequency resources using the tables 6.3.3.2-2 to 6.3.3.2-4 given in TS 38.211~\cite{3gpp_38_211}. After $y_{u, v}(k)$ is mapped to the resource grid, the signal is OFDM modulated, up-converted to the carrier frequency, and transmitted over the channel.
At the receiver, after down-conversion and OFDM demodulation, the frequency domain signal is given by,
\begin{equation}
    \tilde{y}_{u,v}(k) = h(k)X_u(k)e^{\frac{-j2\pi k}{L_{RA}} \left[{C_v - \frac{TA \cdot L_{RA}}{N}}\right]} + w(k), 
    \label{eq: RX_PRACH_with_TA}  
\end{equation}
where $h(k)$ represents the fading channel, $N$ is the OFDM FFT size, and $w(k)$ is the Additive White Gaussian Noise (AWGN). Note that Eq.\eqref{eq: RX_PRACH_with_TA} includes one rotation that uniquely identifies the RAPID (through $C_{v}$ and $u$) and a second rotation that is induced by the propagation delay. In this paper, we assume that $L_{RA} = 139$, $N=4096$, and a sub-carrier spacing of $30$ kHz. Hence a propagation delay of $\frac{4096}{139} = 29.4$ samples (caused by a distance of $71$m)  causes the correlation peak at the receiver to shift by one sample.  In this paper, we evaluate the performance of a small cell with a radius of $850$ meters, which corresponds to $12$ discrete values of TA. 

\begin{table*}[h!]
    \centering
    \caption{Features of various AI/ML approaches.}
    \includegraphics[width=0.7\linewidth]{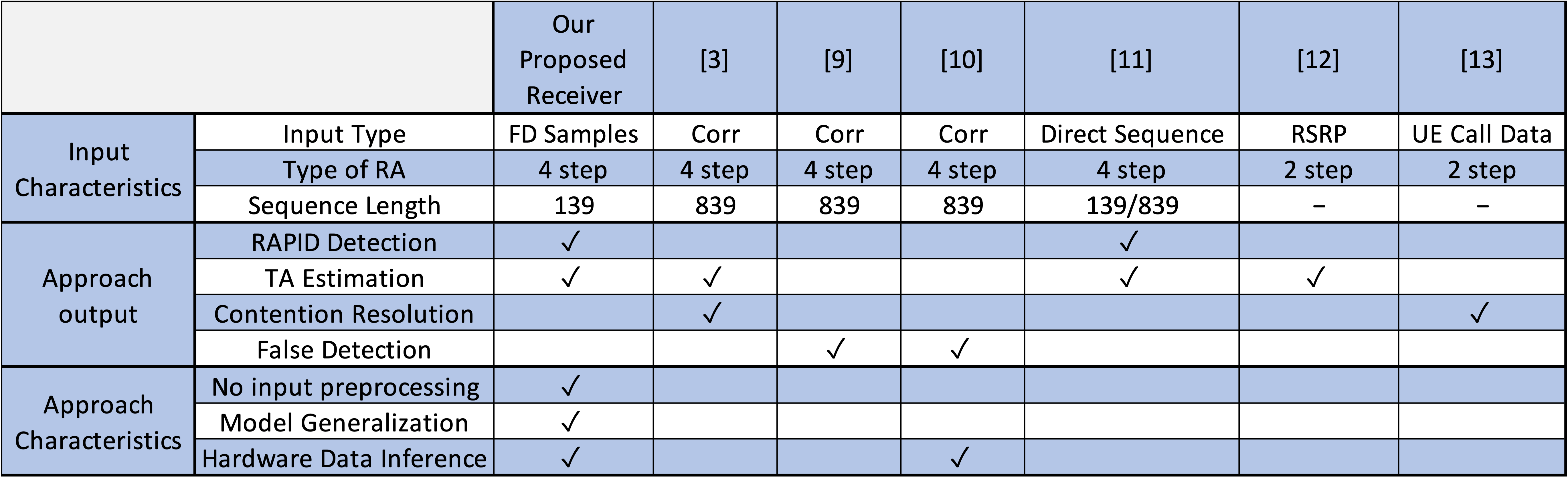}
    \label{tab: ai_ml_features}
\end{table*}

\subsection{Existing receivers for 5G NR PRACH}
The primary goal of a PRACH receiver is to accurately determine the RAPID and the TA. Here, we summarize both the conventional and AI/ML approaches taken by other works in the design of PRACH receivers. 

\subsubsection{Conventional approaches}
Most existing PRACH receivers are correlation-based~\cite{pham2019proposed, kamata2021detection} where the received time domain samples are correlated with all the possible base sequences, and the one that results in the maximum correlation value is identified as the transmitted base sequence. The shift in the correlation peak maps to the RAPID and the TA. If a Zadoff Chu Sequence is correlated with the rotated version of the same sequence, it will result in a strong peak at the shift. If it is correlated with any other sequence, the correlation will be ideally zero (or below a threshold~\cite{hu2012method, threshold_2}).

\begin{figure}[h!]
    \centering
    \includegraphics[width=0.9\linewidth]{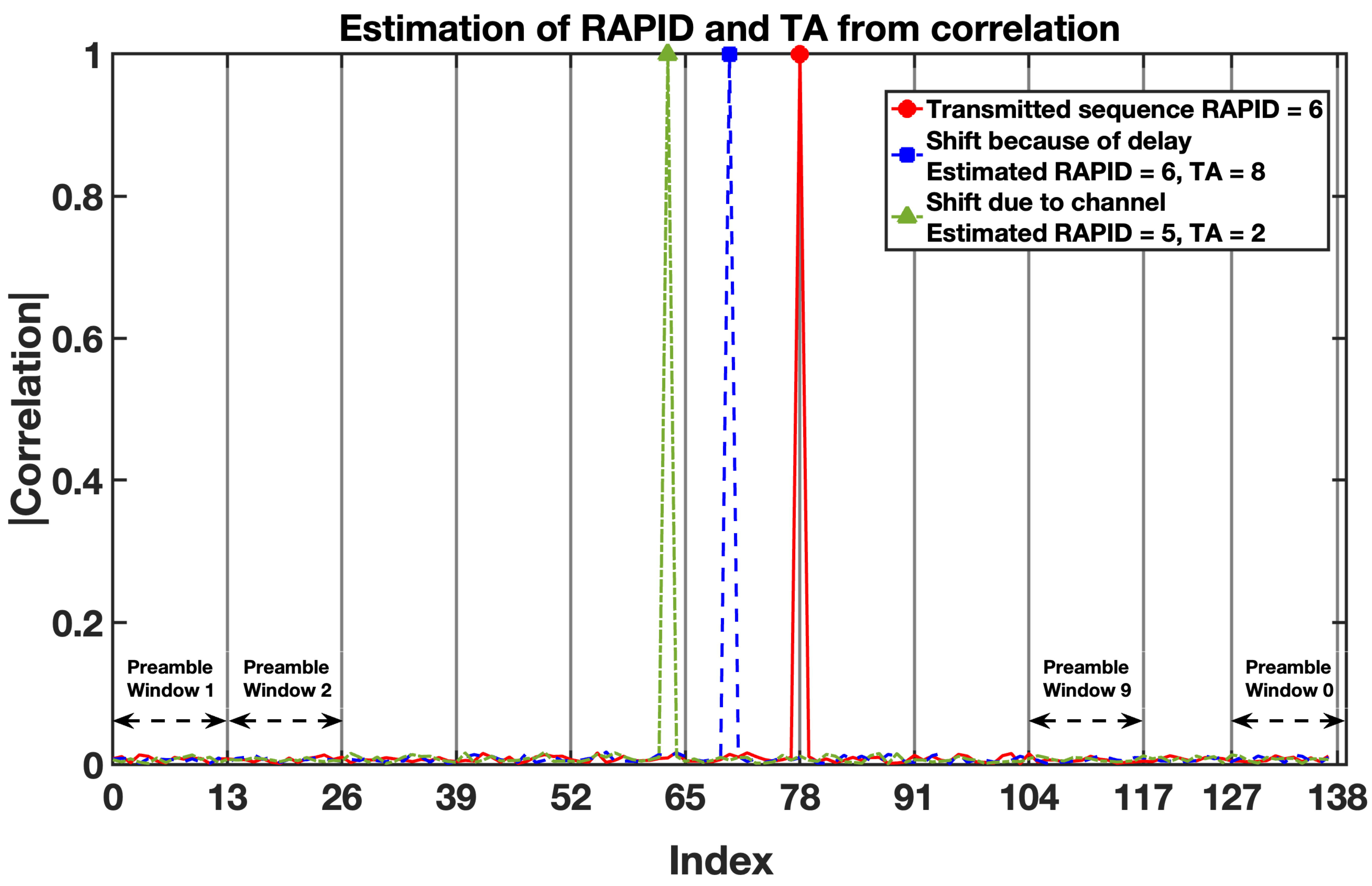}
    \caption{Example correlation plot, showing division into preamble windows and shifting of the peaks due to propagation delay and fading channel effects.}
    \label{fig: corr_windows}
\end{figure}

A common implementation (such as the one used at the IIT Madras 5G testbed~\cite{5gtbiitm}) of a correlation-based PRACH receiver involves dividing the $139$ length correlation output into multiple preamble windows. Each window corresponds to a particular RAPID value and the shift of the peak from the boundary of the window indicates the TA. Fig.~\ref{fig: corr_windows}, shows three example peaks corresponding to three different cases with the same transmitted RAPID value of 6. In the first example, if the transmitted sequence is correlated with the base sequence, the peak occurs at the $78$th position ($C_v = 6\times N_{CS} = 78$) as shown in  Fig.~\ref{fig: corr_windows}. In the second example, due to the propagation delay of $236$ samples, the peak shifts to the $70$th position which is still contained within the same preamble window (Estimated RAPID=6, TA=8 in Fig.~\ref{fig: corr_windows}). In the third example, adverse channel effects along with the propagation delay result in the peak shifting into the next preamble window causing errors in both RAPID detection and TA estimation (Estimated RAPID=5, TA=2 in Fig.~\ref{fig: corr_windows}). 


\subsubsection{AI/ML approaches}

\begin{figure*}[h!]
    \centering
    \includegraphics[width=0.8\textwidth]{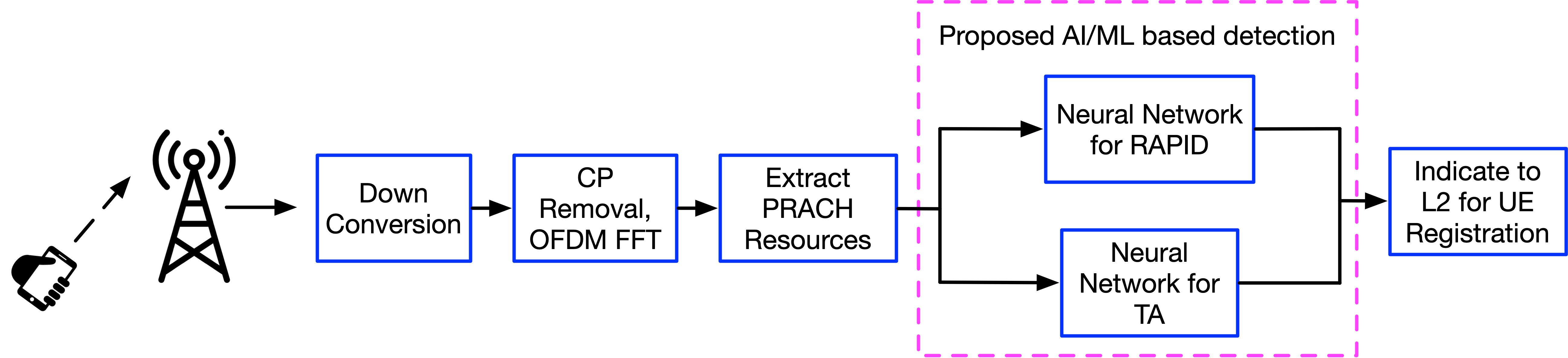}
    \caption{The proposed AI/ML architecture replaces correlation operations with 2 neural networks that can run in parallel.}
    \label{fig: aiml_prach_rx}
\end{figure*}

Any PRACH receiver should be designed for RAPID detection, TA estimation, Collision Detection and Contention Resolution (CD/CR), and false detection. As summarized in Table~\ref{tab: ai_ml_features}, prior work is focussed on one or more of the above outcomes. For ease of understanding the AI/ML landscape for PRACH, we compile and summarize some of the key similarities and differences between our work and some of the prior works in Table~\ref{tab: ai_ml_features}. 

CD and CR are crucial in massively connected IoT scenarios. For such a scenario,~\cite{quek} develops a framework for CD and CR using multiple neural network classifiers. First, classifier models are trained to predict the number of UEs likely to collide. Separate classifiers are designed for TA estimation. AI/ML models such as k-Nearest Neighbors (kNN)~\cite{modina2019machine, zehra2022proactive}, Naive Bayes~\cite{modina2019machine, zehra2022proactive}, and Decision Tree Classification (DTC)~\cite{zehra2022proactive} have shown to improve the True/False classification performance. A further gain is achieved through ensemble learning methods~\cite{zehra2022proactive}. An initial attempt at RAPID detection and TA estimation is made by~\cite{9951528}. However, this work does not consider any impairments such as fading or noise. We note that most of the above methods (1) rely on correlation data for their inputs (2) consider the longer 839-length PRACH sequences and (3) focus on the 4-step RA.

The 2-step RA process, which was introduced later in 5G Release 16, has seen some interest. For 2-step RA, AI/ML-based TA estimation using RSRP values collected at an edge RAN controller is shown in~\cite{9261952}. The trained model is then deployed to different UEs and updated periodically or as needed. A method for collision avoidance for 2-step RA is shown in~\cite{swain2023employing}. Recurrent Neural Networks (RNN) and Long Short Term Memory (LSTM) models are trained with a publicly available dataset containing time instances at which UEs connect to the network. Using the model to predict whether a UE is likely to transmit on the PRACH, the base station preemptively allocates resources to the UE, thus eliminating the need for the UE to even transmit and reducing collisions. 

Since most of the deployed 5G systems around the world use the 4-step RA process, in this paper we focus on RAPID and TA detection for the 4-step RA. We focus on the 5G PRACH A1 format, where a 139-length Zadoff Chu sequence is transmitted by the UE. We observe that while the 139-length sequences are not studied in detail in the current literature there is a significant advantage of the 139-length sequence over the 839-length sequence in that it consumes fewer radio resources. A further distinction of our work is that the input to our proposed AI/ML-based PRACH receiver is directly extracted from the frequency domain resources after performing OFDM-FFT. This is in contrast to other works, which require a correlation of the received samples with the base sequence. Furthermore, we focus on model generalization, another aspect, which to the best of our knowledge, is the first such attempt. The first level of model generalization is with respect to the different simulated channel scenarios (TDLC10, TDLC150, and TDLC300), and the second,
is with respect to the hardware OTA data captured at the IIT Madras 5G Testbed~\cite{5gtbiitm}.

\section{Proposed AI/ML model for the PRACH receiver}
As described in the previous section, both the phase rotations in Eq.~\ref{eq: RX_PRACH_with_TA}, come from discrete sets. Hence RAPID and TA detection can be considered as a classification of one element out of the respective discrete sets.

\subsection{Neural Network Architectures}
\begin{figure}[h]
    \centering
    \includegraphics[width=0.75\linewidth]{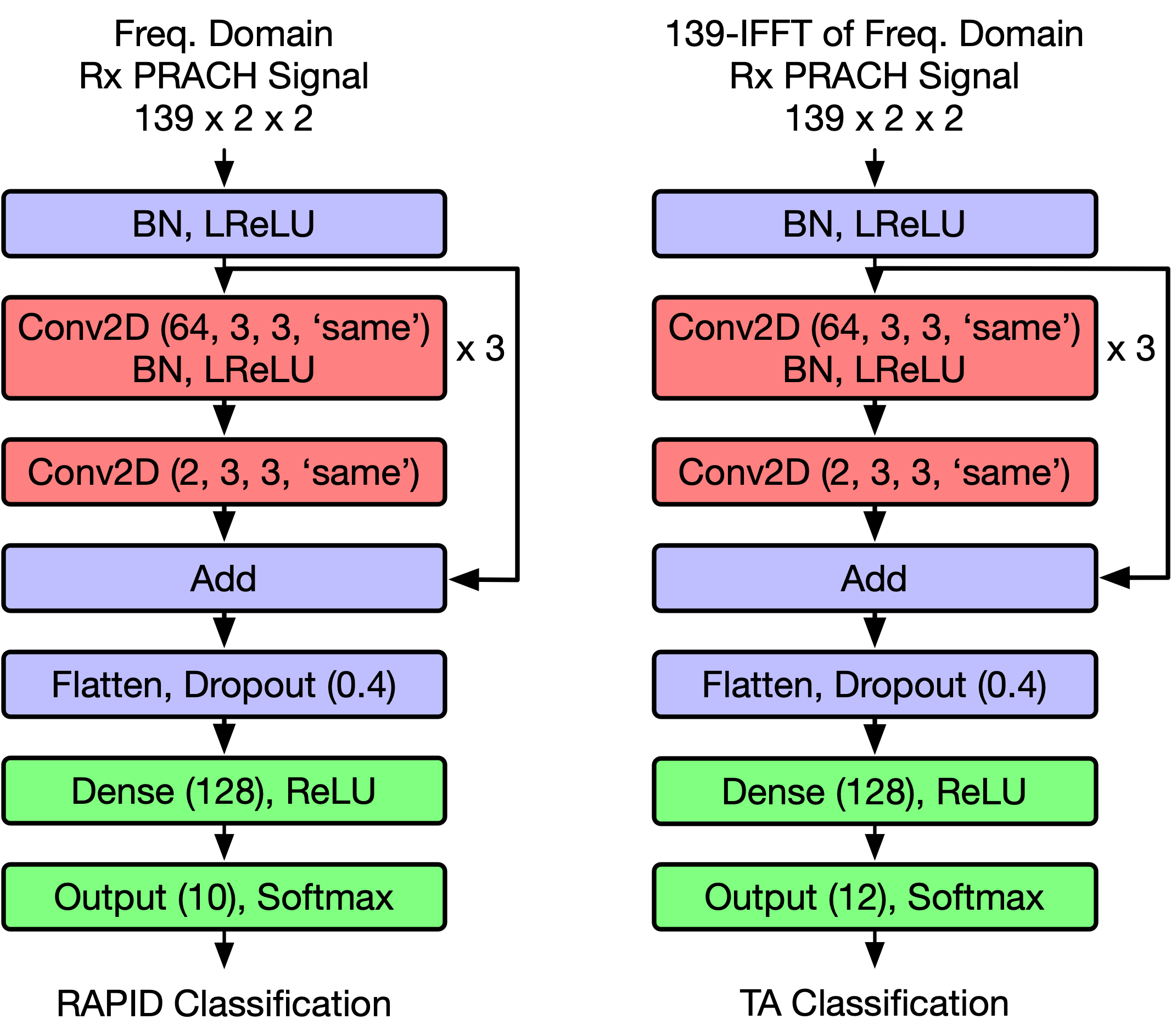}
    \caption{NN Architecture}
    \label{fig: NN_Architecture}
\end{figure}

In this work, we use a Residual Network (ResNet)-like architecture with four convolutional layers to classify both RAPID and TA. ResNets are popular because they use skip connections to overcome vanishing gradients. The input to the RAPID detection model is directly taken from the received frequency domain resources after the OFDM FFT. The input to the TA estimation model is the $139$-point IFFT of the extracted frequency domain data. The input dimension to both networks is $139 \times 2 \times 2$ ($139$ resource elements, $2$ OFDM symbols, and real and imaginary parts of the samples). The input is first passed through a Batch normalization (BN) block followed by a Leaky Rectified Linear Unit (LReLU) activation function. In both models, this is followed by three convolutional layers, each with $64$ filters. Each of these layers is followed by a BN block and a LReLU activation function. The output is passed through another convolutional layer with $2$ filters. The output of the final convolutional layer is added to the output of the initial BN + LReLU output. The sum is then flattened and passed to the output via a dense layer of size $128$, with ReLU activation. After the flatten operation, we use dropout as a regularizer with a probability of $0.4$. The RAPID output layer is of size $10$ and the TA output layer is of size $12$. Both output layers use the softmax activation function. Stochastic Gradient Descent (SGD) optimizer with a learning rate of $10^{-2}$ is used in case of RAPID detection and an Adam optimizer with a learning rate of $10^{-3}$ is used for TA estimation. The detailed architecture of both the neural networks is shown in Fig.~\ref{fig: NN_Architecture}.

\subsection{Dataset Generation}
Two types of data are considered in this paper - simulated data and Over-the-Air (OTA) data captured on testbed hardware. 

\subsubsection{Simulated datasets}
These datasets are generated using the MATLAB 5G Toolbox. We consider an SNR range of $-20$ dB to $+20$ dB in the steps of $5$ dB. For each SNR, $100000$ data instances are generated. Each data instance is generated by randomly choosing a RAPID and applying it to the base sequence. The corresponding OFDM-modulated signal is transmitted over a TDLC channel and the received samples are extracted. These are the inputs to the models. The labels are the chosen RAPID values. To simulate a propagation delay (timing advance), a random number of zero samples are added before transmitting the signal over the channel. This translates to a phase shift in the frequency domain at the receiver. In this case, the ground truth TA label is calculated as the floor of the ratio of the propagation delay to 29.4.

\subsubsection{Over-the-Air (OTA) Hardware captured datasets}
\begin{figure}[ht]
\centering
\includegraphics[width=0.3\textwidth]{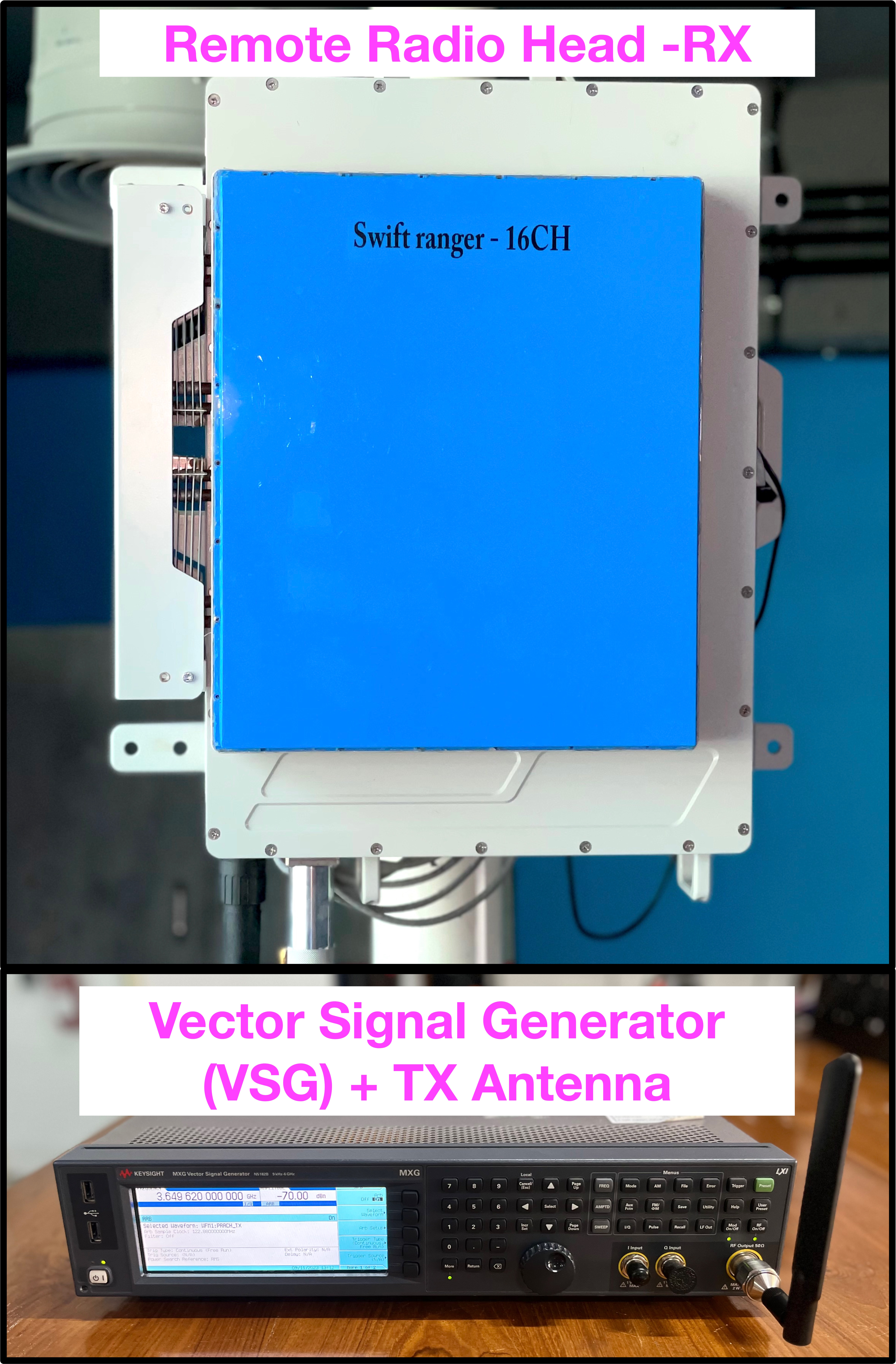}
\caption{IIT-Madras 5G testbed setup with the Remote Radio Head used as a receiver and the VSG used as a transmitter.}
\label{fig: hw_setup}
\end{figure}

Validation of AI/ML models with data collected from deployed base station provides insights into the robustness of the model with respect to the effects of hardware elements like DACs, ADCs, up/down converters, amplifiers, etc. We captured the data on the hardware deployed at the state-of-the-art 5G testbed at IIT Madras~\cite{5gtbiitm} to analyze model performance. The captures help incorporate real-time channel conditions and hardware impairments that may/may not be included in the training dataset. The setup (as shown in Fig.~\ref{fig: hw_setup}) consists of an N5182B Vector Signal Generator (VSG) for transmitting the  5G signal at a center frequency of $3.64962$ GHz (sub-6 GHz raster in the n78 band). The commercial omnidirectional wideband monopole antenna transmits signals originating from the VSG.

A  multi-channel receiver front end connected to a dual-polarized antenna at the base station receives the signals over the air (for this paper, we utilize only one antenna and one transceiver chain). Other receiver components include an in-house Low Noise Amplifier (LNA) with 60dB gain at the receiver front end and an ADRV 9009 RF transceiver. We place the transmitter about one meter away from the receiver to emulate one of the real-time wireless scenarios. The PRACH signal is transmitted from the VSG through an antenna, over the air, and then received at the base station's antenna, amplified at the LNA, followed by the transceiver. The signal out of the 16-bit ADC is then collected and used for testing. A total of $56000$ OTA data instances are captured with the hardware.

\subsection{Training and testing}
In this paper, we consider $3$ cases for training and testing, as described below. In all cases, we use a $75-25$\% train-test split. Of the training dataset, $30$\% of the data is for validation. In all cases, training is done with a combination of data corresponding to $8$ SNRs ($-15$dB to $20$dB in steps of $5$dB). 

\subsubsection{Training and testing on same channel scenario (Simulated Datasets)}
We train both the RAPID and TA classifier models with datasets that include signals received over TDLC150 channels (Training dataset size = $600000$). Testing is performed on data of the same channel scenario. We repeat the process for datasets that include signals received over TDLC300 channels (Training dataset size = $600000$). These experiments highlight the best-case performance for the simulated datasets. 
\subsubsection{Training and testing on different channel scenarios (Simulated Datasets)}
We generate datasets with signals received over TDLC10, TDLC150, and TDLC300 channels. First, we train the model with a combined dataset (Training dataset size = $1800000$) containing signals received over all 3 channel conditions. We then compare this training with signals received over only TDLC300 channels (Training dataset size = 600000). Testing is done with TDLC150 data. The experiments help analyze the generalization performance of the model across channels of different delay spread values. 
\subsubsection{Training and testing on OTA Hardware datasets}
In the case of OTA hardware captures, we look at three scenarios. First, the OTA test data is fed to the model trained on a mixed dataset containing signals received on TDLC10, TDLC150, and TDLC300 channels. Second, the OTA test data is fed to a model trained on a combination of the mixed simulation data as well as some OTA data  (Training dataset size = $1800000 + 42000$). Lastly, the OTA test data is fed to a model trained purely with OTA data (Training dataset size = $42000$). These experiments help determine the generalization capability of the model across simulation and hardware data. 

\section{Results}
\begin{figure}[h!]
	\centering
	\includegraphics[width=0.8\linewidth]{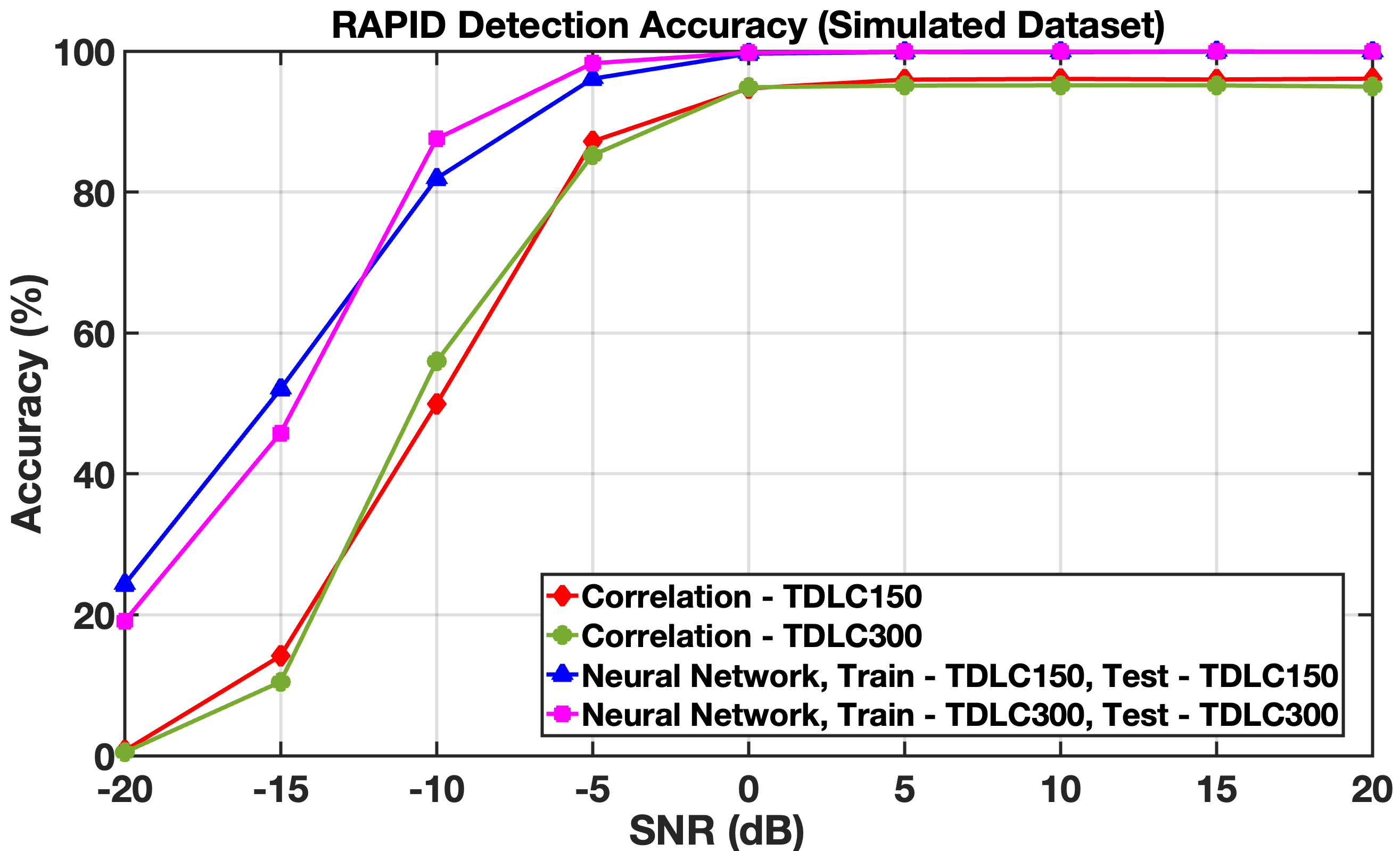}
	\caption{RAPID Detection Accuracy of the NN compared with correlation-based detection. Testing is performed on the same channel scenario as training}
	\label{fig:1_rapid_det_acc_vs_snr_sim}
\end{figure}

\begin{figure}[h!]
	\centering
	\includegraphics[width=0.8\linewidth]{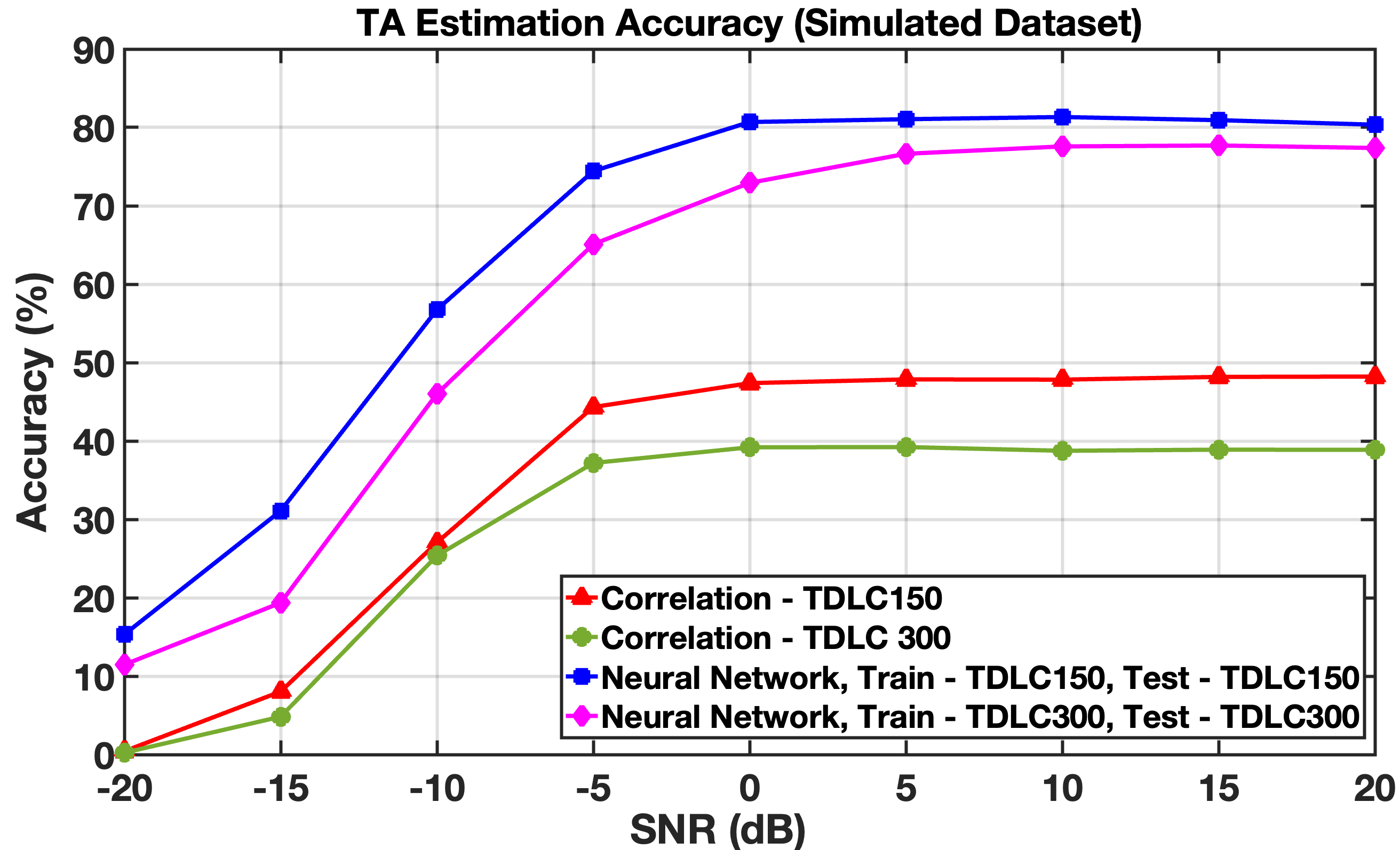}
	\caption{Timing Advance Estimation Accuracy of the NN compared with correlation-based estimation. Testing is performed on the same channel scenario as training}
	\label{fig:2_ta_est_acc_vs_snr_sim}
\end{figure}

This section shows the performance analysis of the proposed neural network models and compares it with that of conventional correlation-based approaches. 
Fig.~\ref{fig:1_rapid_det_acc_vs_snr_sim} and Fig.~\ref{fig:2_ta_est_acc_vs_snr_sim} show the accuracy of RAPID detection and TA Estimation respectively. We can see that NN clearly outperforms the correlation-based approach in both cases. In the lower SNR range, the correlation-based approach is impaired by both fading and noise. In the higher SNR range, the noise reduces but the fading remains, leading to a saturation in the accuracy.  This is because in channels such as TDLC150 and TDLC300, the number of taps is high and each tap, other than the first tap, causes a different shift in the correlation peak. If the power of the first tap is significant compared to other taps, the other correlation peaks will be insignificant. In the case of TDLC150 and TDLC300 channels, we observed that the other taps have significant power and the combined influence of those taps on the correlation peak leads to incorrect detections. One shift in the detected peak from the correct location leads to a shift in the estimated TA by one. For $N_{CS} $= 13, it takes 13 shifts of the TA to make the detected RAPID incorrect by 1. Hence RAPID detection is more robust than TA estimation. 

\begin{figure}[h!]
    \centering
    \includegraphics[width=0.8\linewidth]{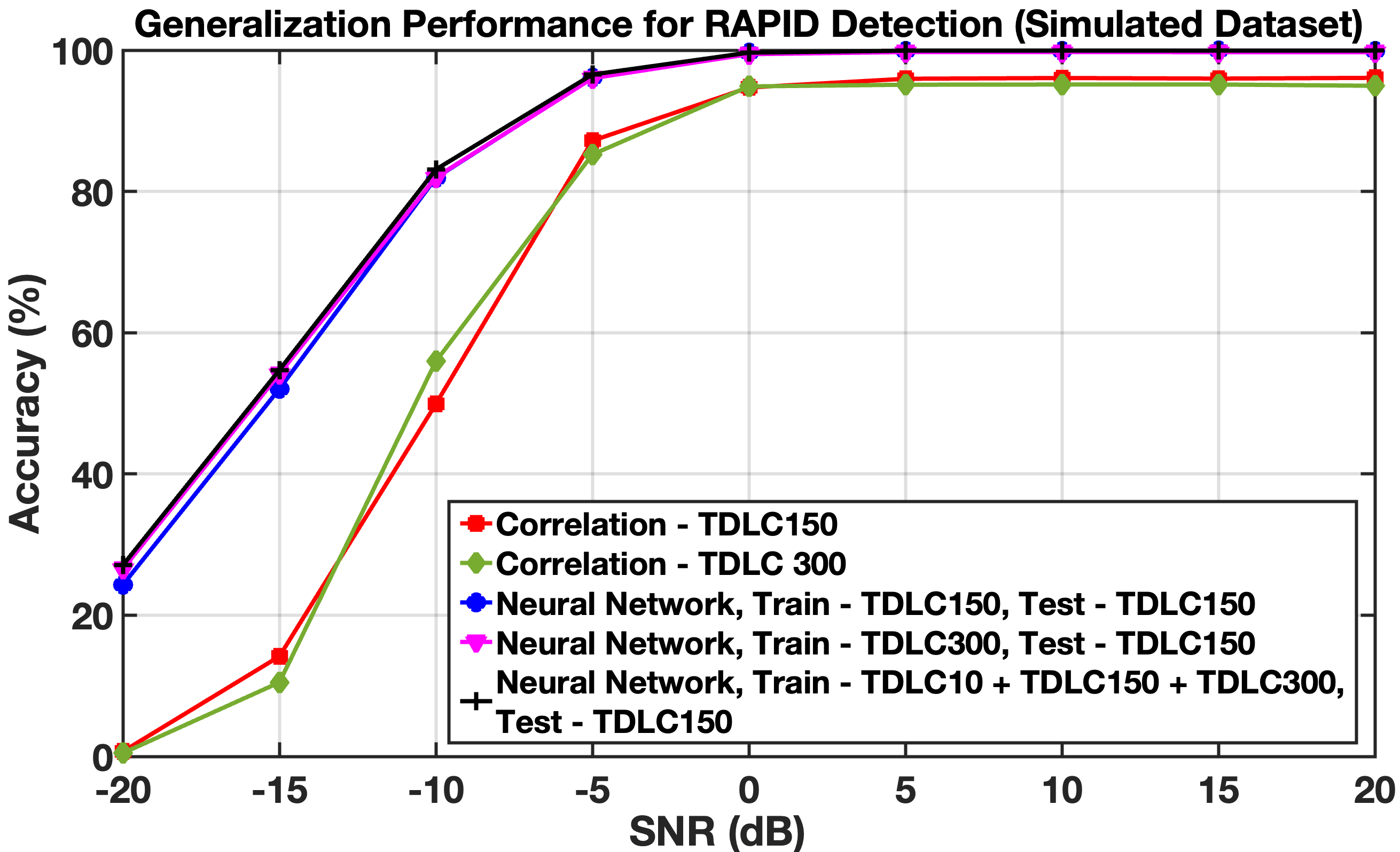}
    \caption{Generalization performance of the NN-based RAPID detection across channel scenarios}
    \label{fig:3_gen_rapid_acc_vs_snr_sim}
\end{figure}

\begin{figure}[h!]
    \centering
    \includegraphics[width=0.8\linewidth]{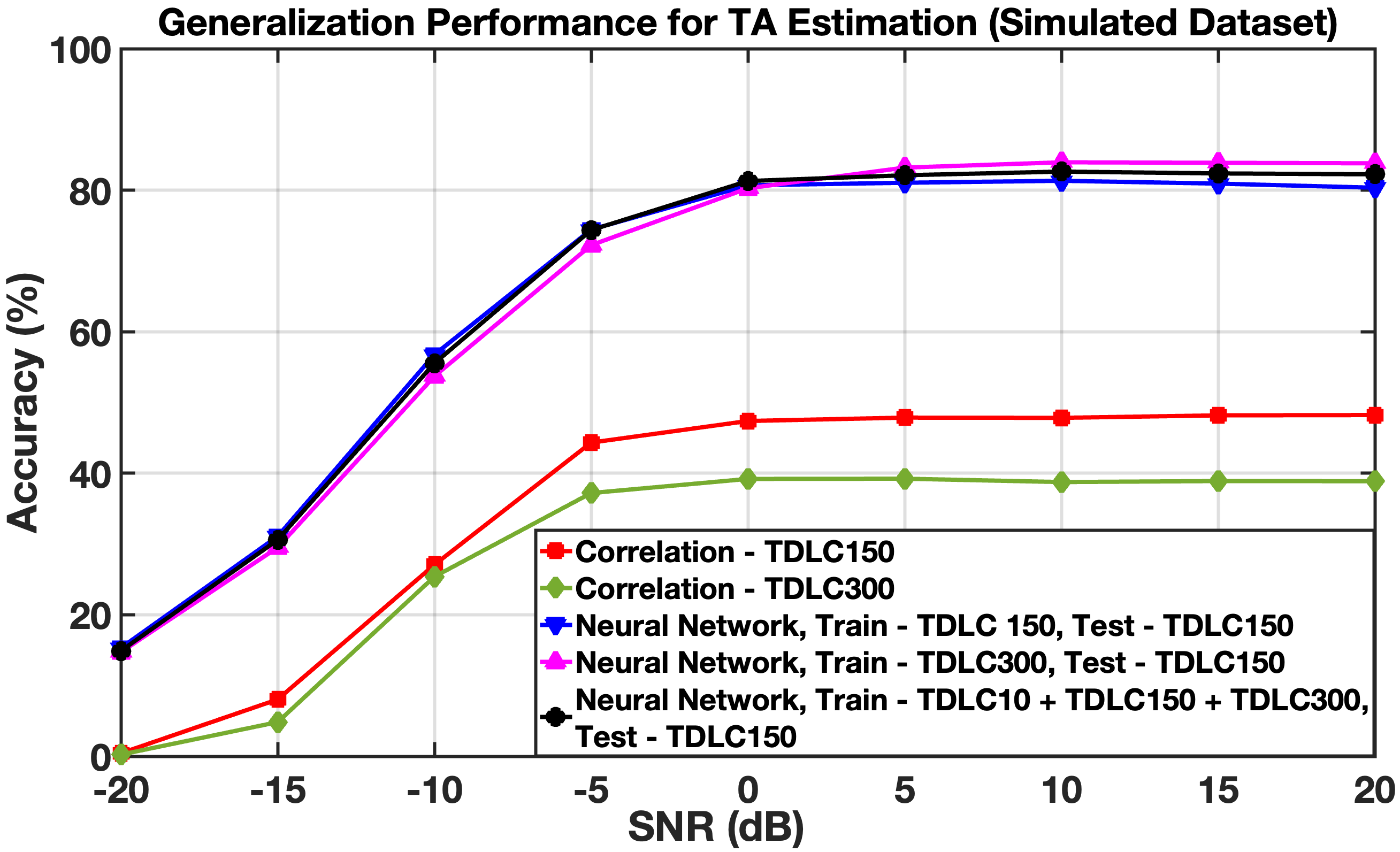}
    \caption{Generalization performance of the NN-based Timing Advance estimation across channel scenarios.}
    \label{fig:4_gen_ta_est_vs_snr_sim}
\end{figure}

Fig.~\ref{fig:3_gen_rapid_acc_vs_snr_sim} and Fig.~\ref{fig:4_gen_ta_est_vs_snr_sim} focus on the generalization of the model in RAPID detection and TA estimation. We observe that if we train the NN with a mixed dataset of signals received over various channel scenarios, testing with one subset of the dataset does not result in a drop in performance. Training with mixed datasets, however, can be memory and time-prohibitive. Consequently, we train the model with signals received over a more complex TDLC300 channel and test with data from a less complex channel such as TDLC150. We note that there is no significant drop in performance compared to the mixed training case. 
    
\begin{figure}[h!]
    \centering
    \includegraphics[width = 0.8\linewidth]{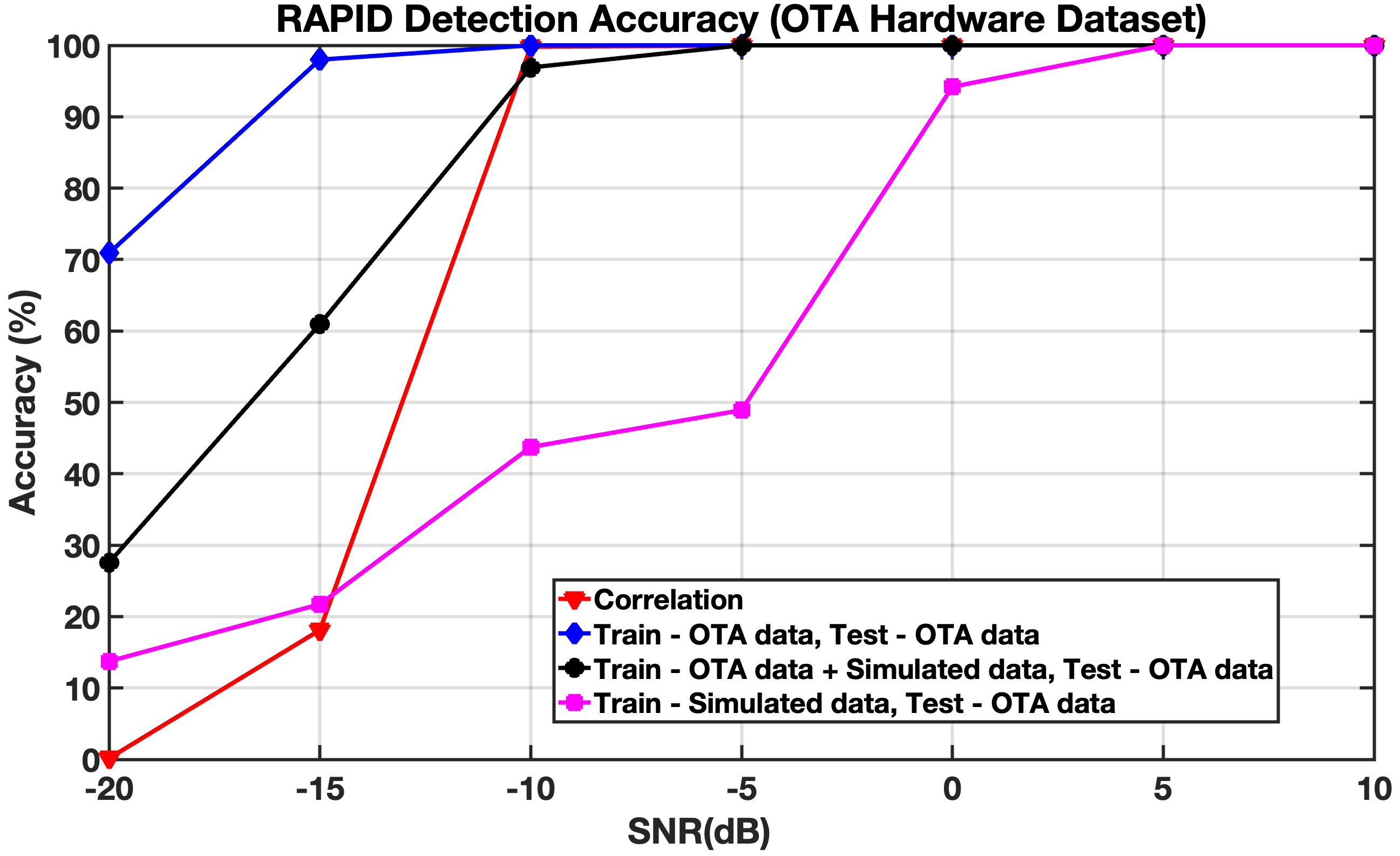}
    \caption{RAPID detection accuracy vs SNR for OTA hardware datasets}   \label{fig:5_rapid_acc_vs_snr_ota}
\end{figure}

Lastly, Fig.~\ref{fig:5_rapid_acc_vs_snr_ota} shows the RAPID detection accuracy vs SNR for the OTA captured data. When the neural network is trained with simulation data alone, it is unable to capture the impairments that are caused by the hardware elements on the receiver path such as ADC, filters, Low Noise Amplifiers (LNA), etc. To include these impairments, we first train with a pure hardware dataset. This is the best-case scenario, but it is impractical due to the complex nature of the capturing process. A more practical training approach is to combine hardware data with simulated data. Currently, we are limited by the amount of captures that can be taken from the testbed. With a sufficiently large data collection, the performance of combined training will approach pure hardware training. Note that these hardware captures are taken in a lab environment where the transmitter and receiver are one meter apart, resulting in a Line-of-sight scenario. Hence, correlation-based decoding also performs well at SNRs greater than -10 dB.

\section{Conclusion}
A robust PRACH receiver helps users connect to the network seamlessly. In this work, we have built one such receiver that is implemented using two parallel and independent Neural Networks, one for RAPID detection and another for TA estimation. The NNs take frequency domain samples directly extracted from the OFDM resource grid as input rather than the output of operations such as correlation. Results show that the NN-based PRACH receiver outperforms traditional correlation-based receivers for a single base sequence. This indicates that our AI/ML-based receiver decodes the PRACH signal with a higher probability, which will lead to a reduction in PRACH retransmissions. We have also shown initial promising indications that the NN generalizes well across channel scenarios of various delay spreads and Over-the-Air data captured on a deployed base station. Future work will incorporate more base sequences, more variations in channel scenarios (simulated and OTA), and multi-user collision detection.

\section*{Acknowledgment}
The authors would like to thank the Department of Telecommunications (DOT), India,  for funding the 5G Testbed project and the Ministry of Electronics and Information Technology (MeitY) for funding this work through the project "Next Generation Wireless Research and Standardization on 5G and Beyond".


\bibliographystyle{IEEEtran}
\bibliography{bibfile}
\end{document}